\newcommand{\CLASSINPUTtoptextmargin}{0.75in}%
\newcommand{\CLASSINPUTbottomtextmargin}{1in}%
\newcommand{\CLASSINPUTinnersidemargin}{0.63in}%
\newcommand{\CLASSINPUToutersidemargin}{0.63in}%
\documentclass[conference,10pt,letterpaper]{IEEEtran}% 
\usepackage{amsmath}% for double integral symbol in this template
\usepackage{mathtools} % Provides \mathrlap
\usepackage{graphicx}% for figures
\usepackage{grfext} 
\usepackage{xcolor}

\usepackage{multirow}% to allow multiple-row elements in tabular environment
\usepackage[none]{hyphenat}% turn off hyphenation to make text extraction and indexing easier
\usepackage{float}% better control of floating figures and tables
\usepackage{subfig}% for subfigures within figures
\usepackage{t1enc}% allows access to various special characters
\usepackage{times}% change font to Nimbus Roman (based on Times Roman)
\usepackage[numbers, square, comma, sort&compress]{natbib}
\usepackage{comment}
\makeatletter
\def\ps@IEEEtitlepagestyle{%
  \def\@oddfoot{\mycopyrightnotice}%
  \def\@evenfoot{}%
}
\def\mycopyrightnotice{%
  { \centering\normalsize
  \hspace{+6cm}
  979-8-3315-0812-8/25\$31.00 \textcopyright2025 IEEE\hfill}%
  \gdef\mycopyrightnotice{}% just in case
}

\makeatletter

\def\@RFICauthorblockNAMEstyle{\normalfont\RFICauthorsize}
\def\@RFICauthorblockAFFILstyle{\normalfont\RFICaffilsize}
\def\@RFICauthorblockEMAILstyle{\normalfont\RFICaffilsize}
\def\RFICauthorblockNAME#1{%
\relax\@RFICauthorblockNAMEstyle%
#1%
}%
\def\RFICauthorblockAFFIL#1{%
\relax\@RFICauthorblockAFFILstyle%
\vskip\@IEEEauthorblockAtopspace
#1%
}%
\def\RFICauthorblockEMAIL#1{%
\relax\@RFICauthorblockEMAILstyle%
\vskip\@IEEEauthorblockAtopspace
#1%
}%
\newcommand{\RFICauthor}[1]{%
\ifIsBlindReviewVersion%
\author{\phantom{\parbox{\textwidth}{\center\relax#1}}}%
\else%
\author{\parbox{\textwidth}{\center\relax#1}}%
\fi%
}%
\newif\ifIsBlindReviewVersion

\def\RFICthispaperforfinalpublication{\IsBlindReviewVersionfalse}
\def\@maketitle{\newpage
\bgroup\par\addvspace{0.5\baselineskip}\centering%
\ifCLASSOPTIONtechnote% technotes
   {\bfseries\large\@IEEEcompsoconly{\sffamily}\@title\par}\vskip 1.3em{\lineskip .5em\@IEEEcompsoconly{\sffamily}\@author
   \@IEEEspecialpapernotice\par{\@IEEEcompsoconly{\vskip 1.5em\relax
   \@IEEEtitleabstractindextextbox{\@IEEEtitleabstractindextext}\par
   \hfill\@IEEEcompsocdiamondline\hfill\hbox{}\par}}}\relax
\else% not a technote
   \vskip0.2em{\RFICtitlesize\ifCLASSOPTIONtransmag\bfseries\LARGE\fi\@IEEEcompsoconly{\sffamily}\@IEEEcompsocconfonly{\normalfont\normalsize\vskip 2\@IEEEnormalsizeunitybaselineskip
   \bfseries\Large}\@title\par}\vskip1.0em\par% CAUSAL PRODUCTIONS change on this line
   \ifCLASSOPTIONconference%
      {\@IEEEspecialpapernotice\mbox{}\vskip\@IEEEauthorblockconfadjspace%
       \mbox{}\hfill\begin{@IEEEauthorhalign}\@author\end{@IEEEauthorhalign}\hfill\mbox{}\par}\relax
   \else% peerreviewca, peerreview or journal
      \ifCLASSOPTIONpeerreviewca
         {\@IEEEcompsoconly{\sffamily}\@IEEEspecialpapernotice\mbox{}\vskip\@IEEEauthorblockconfadjspace%
          \mbox{}\hfill\begin{@IEEEauthorhalign}\@author\end{@IEEEauthorhalign}\hfill\mbox{}\par
          {\@IEEEcompsoconly{\vskip 1.5em\relax
           \@IEEEtitleabstractindextextbox{\@IEEEtitleabstractindextext}\par\hfill
           \@IEEEcompsocdiamondline\hfill\hbox{}\par}}}\relax
      \else% journal, peerreview or transmag
         \ifCLASSOPTIONtransmag
           {\@IEEEspecialpapernotice\mbox{}\vskip\@IEEEauthorblockconfadjspace%
            \mbox{}\hfill\begin{@IEEEauthorhalign}\@author\end{@IEEEauthorhalign}\hfill\mbox{}\par
           {\vspace{0.5\baselineskip}\relax\@IEEEtitleabstractindextextbox{\@IEEEtitleabstractindextext}\vspace{-1\baselineskip}\par}}\relax
         \else% journal or peerreview
           {\lineskip.5em\@IEEEcompsoconly{\sffamily}\sublargesize\@author\@IEEEspecialpapernotice\par
           {\@IEEEcompsoconly{\vskip 1.5em\relax
            \@IEEEtitleabstractindextextbox{\@IEEEtitleabstractindextext}\par\hfill
            \@IEEEcompsocdiamondline\hfill\hbox{}\par}}}\relax
         \fi
      \fi
   \fi
\fi\par\addvspace{0.0\baselineskip}\egroup}% CAUSAL PRODUCTIONS change on this line, reduce the vspace from 0.5\baselineskip to 0.0

\def\RFICtitlesize{\@setfontsize{\RFICtitlesize}{18}{21pt}}% CAUSAL PRODUCTIONS change on this line
\def\RFICauthorsize{\@setfontsize{\RFICauthorsize}{12}{13pt}}% CAUSAL PRODUCTIONS change on this line
\def\RFICaffilsize{\@setfontsize{\RFICaffilsize}{12}{13pt}}% CAUSAL PRODUCTIONS change on this line
\def\RFICcaptionsize{\@setfontsize{\RFICcaptionsize}{8}{9pt}}% CAUSAL PRODUCTIONS change on this line
\def\RFICbibsize{\@setfontsize{\RFICbibsize}{8}{9pt}}% CAUSAL PRODUCTIONS change on this line

\def\@IEEEauthorblockNstyle{\RFICauthorsize\@IEEEcompsocnotconfonly{\sffamily}\@IEEEcompsocconfonly{\large}}%CAUSAL PRODUCTIONS removed sublargesize to get correct RFICauthorsize
\def\@IEEEauthorblockAstyle{\RFICaffilsize\@IEEEcompsocnotconfonly{\sffamily}\@IEEEcompsocconfonly{\itshape}\@IEEEcompsocconfonly{\large}}%CAUSAL PRODUCTIONS removed normalsize to get correct RFICaffilsize
\def\@IEEEauthordefaulttextstyle{\RFICauthorsize\@IEEEcompsocnotconfonly{\sffamily}\sublargesize}%CAUSAL PRODUCTIONS

\def\thebibliography#1{\section*{\refname}%
    \addcontentsline{toc}{section}{\refname}%
    \RFICbibsize\@IEEEcompsocconfonly{\small}\vskip 0.3\baselineskip plus 0.1\baselineskip minus 0.1\baselineskip% CAUSAL PRODUCTIONS change on this line
    \list{\@biblabel{\@arabic\c@enumiv}}%
    {\settowidth\labelwidth{\@biblabel{#1}}%
    \leftmargin\labelwidth
    \advance\leftmargin\labelsep\relax
    \itemsep \IEEEbibitemsep\relax
    \usecounter{enumiv}%
    \let\p@enumiv\@empty
    \renewcommand\theenumiv{\@arabic\c@enumiv}}%
    \let\@IEEElatexbibitem\bibitem%
    \def\bibitem{\@IEEEbibitemprefix\@IEEElatexbibitem}%
\def\newblock{\hskip .11em plus .33em minus .07em}%
\ifCLASSOPTIONtechnote\sloppy\clubpenalty4000\widowpenalty4000\interlinepenalty100%
\else\sloppy\clubpenalty4000\widowpenalty4000\interlinepenalty500\fi%
    \sfcode`\.=1000\relax}

\long\def\@makecaption#1#2{%
\ifx\@captype\@IEEEtablestring%
\par\@IEEEtabletopskipstrut% strut used to align table caption with facing column
\else
\@IEEEfigurecaptionsepspace
\fi
\setbox\@tempboxa\hbox{\normalfont\RFICcaptionsize {#1.}\nobreakspace\nobreakspace #2}%
\ifdim \wd\@tempboxa >\hsize%
\setbox\@tempboxa\hbox{\normalfont\RFICcaptionsize {#1.}\nobreakspace\nobreakspace}%
\parbox[t]{\hsize}{\normalfont\RFICcaptionsize\noindent\unhbox\@tempboxa#2}%
\else
\ifCLASSOPTIONconference \hbox to\hsize{\normalfont\RFICcaptionsize\hfil\box\@tempboxa\hfil}%
\else \hbox to\hsize{\normalfont\RFICcaptionsize\box\@tempboxa\hfil}%
\fi\fi
\ifx\@captype\@IEEEtablestring%
\@IEEEtablecaptionsepspace
\else
\fi}

\newlength\tablecaptiontotableskip
\newlength\figuretocaptionskip
\def\@IEEEfigurecaptionsepspace{\vskip\figuretocaptionskip\relax}%
\def\@IEEEtablecaptionsepspace{\vskip\tablecaptiontotableskip\relax}%

\def\abstract{\normalfont%
\@IEEEabskeysecsize\bfseries\textit{\abstractname}\,\bfseries\textit{---}\,%
\@IEEEgobbleleadPARNLSP}%

\def\IEEEkeywords{\normalfont%
\@IEEEabskeysecsize\bfseries\textit{\IEEEkeywordsname}\,\bfseries\textit{---}\,%
\@IEEEgobbleleadPARNLSP}%
\def\endIEEEkeywords{\relax\vspace{0.67ex}%
\par\if@twocolumn\else\endquotation\fi%
\normalsize\normalfont}%

\DeclareRobustCommand*{\RFICauthorrefmark}[1]{\raisebox{0pt}[0pt][0pt]{\textsuperscript{\footnotesize{#1}}}}%
\def\@IEEEauthorblockNtopspace{0ex}
\def\@IEEEauthorblockAtopspace{1mm}
\def\tablename{Table}% RFIC: was TABLE in IEEETran, but we are now using capitalized caption text not smallcaps
\def\thetable{\arabic{table}}% RFIC: Tables are numbered using arabic not roman
\def\IEEEkeywordsname{Keywords}% use Keywords instead of Index Terms
\def\subsubsection{\@startsection{subsubsection}{3}{\z@}{1.5ex plus 1.5ex minus 0.5ex}%
{0.7ex plus .5ex minus 0ex}{\normalfont\normalsize\itshape}}%
\def\@seccntformat#1{\csname the#1dis\endcsname\relax}% moved the spacer \hskip 0.5em to individual handlers below
\def\thesectiondis{\thesection.\hskip 0.5em}%					I.	% CAUSAL PRODUCTIONS: moved the hskip spacer from \@seccntformat to here
\def\thesubsectiondis{{\hbox to\parindent{\Alph{subsection}.}}}%		B.	% CAUSAL PRODUCTIONS: indent the subsection name to match paragraph indent
\def\thesubsubsectiondis{{\hbox to \parindent{\arabic{subsubsection})}}}%	3)	% CAUSAL PRODUCTIONS: indent the subsubsection name to match paragraph indent
\def\theparagraphdis{{\hbox to \parindent{\alph{paragraph})}}}%			d)	% CAUSAL PRODUCTIONS: indent the subsubsubsection name to match paragraph indent
\IEEEilabelindentA \parindent
\IEEEilabelindent \IEEEilabelindentA
\IEEEelabelindent \parindent
\IEEEdlabelindent \parindent
\IEEElabelindent \parindent
\newlength\@RFICparindent
\newcommand\RFICdisplayacksection[1]{%
\ifIsBlindReviewVersion%
\noindent\phantom{\parbox[t]{\columnwidth}{\normalbaselines\setlength{\parindent}{\@RFICparindent}{#1}\strut}}%\RFICacktext
\else%
\noindent\parbox[t]{\columnwidth}{\normalbaselines\setlength{\parindent}{\@RFICparindent}{#1}\strut}%
\fi%
}%

\makeatother

\usepackage{blindtext}

\begin{document}
%%%%%%%%%%%%%%%%%%%%%%%%%%%%%%%%%%%%%%%%%%%%%%%%%%%%%%%%%%%%%%%%%%%%%%%%%%%%%
% We use \raggedbottom to avoid latex adding vertical space around headings.
% This gives a better idea to the author about how much white space remains
% as the page limit is approached.
\raggedbottom
%
%%%%%%%%%%%%%%%%%%%%%%%%%%%%%%%%%%%%%%%%%%%%%%%%%%%%%%%%%%%%%%%%%%%%%%%%%%%%%
% PAPER TITLE AND AUTHOR BLOCK
%
% The paper title can use linebreaks \\ within to get better formatting if desired.
%
\title{On the Performance of Multi-Wavelength Underwater Optical Channels in the Presence of Optical Turbulence}
%
% Next we define the author names and affiliations.
% Author names are listed using \RFICauthorblockNAME{} with comma separators between names.
% Affiliations are listed using \RFICauthorblockAFFIL{} with \\ separators between affiliations.
% Email addresses are listed using \RFICauthorblockEMAIL{} with comma separators between emails.
% See below for examples of each of these.
%
% Symbols marking author-affiliation relations are output using \RFICauthorrefmark{}.
%
% Next we typeset the authorblock either as visible text, or as an empty
% box of the same size, based on the value of the Blind Review Flag.
% Note that the Blind Review Flag also determines whether the Acknowledgements
% section is visible or invisible.
% To set the flag to Blind Review mode, simply uncomment the next line
%\RFICthispaperforblindreview
% or to set the flag to Final Paper mode (with author block visible) then
% simply uncomment the next line:
\RFICthispaperforfinalpublication
\RFICauthor{%
\RFICauthorblockNAME{% Author Names
Shideh Tayebnaimi\RFICauthorrefmark{\#1} and
Kamran Kiasaleh\RFICauthorrefmark{\#2}
}% end of \RFICauthorblockNAME
\\%
\RFICauthorblockAFFIL{% Author Affiliations
\RFICauthorrefmark{\#}Optical Communications Laboratory, University of Texas at Dallas, USA
}% end of \RFICauthorblockAFFIL
\\%
\RFICauthorblockEMAIL{% Author Emails
\RFICauthorrefmark{1}Shideh.Tayebnaimi@utdallas.edu, \RFICauthorrefmark{2}kamran@utdallas.edu
}% end of \RFICauthorblockEMAIL
}% end of \RFICauthor
%
% Next we make the title/author block using the information defined above.
\maketitle
%
%%%%%%%%%%%%%%%%%%%%%%%%%%%%%%%%%%%%%%%%%%%%%%%%%%%%%%%%%%%%%%%%%%%%%%%%%%%%%
% ABSTRACT paragraph.
% in the abstract paragraph.
%
\begin{abstract}
This paper presents an analysis of the performance of a Gaussian optical beam as it propagates through the turbulent underwater optical channel (UWOC) under the weak turbulence regime, where optical signal experiences significant fading and scattering, all of which can severely degrade communication quality. It is assumed that on-off keying (OOK) modulation with direct detection is utilized to establish a duplex communication link. A multi-wavelength beam approach is implemented to enhance the performance by leveraging the distinct propagation characteristics of different wavelengths. Performance is established in terms of the probability of fade, number of fade per second, mean fade duration, and mean time between fades. The use of multi-wavelength beam is shown to enhance performance by a sizable margin. 
\end{abstract}
%
%%%%%%%%%%%%%%%%%%%%%%%%%%%%%%%%%%%%%%%%%%%%%%%%%%%%%%%%%%%%%%%%%%%%%%%%%%%%%
% THE REST OF THE PAPER follows.
%

\section{Introduction}

The need for underwater communication has grown significantly in recent years, driven by various applications such as environmental monitoring, scientific data collection, submerged platform communication, and maritime archaeology. These diverse use cases motivate studies of reliable wireless communication systems capable of operating effectively in challenging underwater environments \cite{korotkovalightinturbulentocean}. As a result, there has been considerable interest in exploring the potential of underwater wireless optical communications (UWOC), which offer high data rates, low latency, and efficient transmission for these applications. UWOC technologies promise to play a pivotal role in the improvement of various industries, ranging from oceanography and underwater exploration to military and commercial maritime operations.

However, UWOC systems are typically limited in link length, usually spanning only tens of meters, due to turbulence, scattering, and absorption in the underwater medium. These limitations arise from unpredictable and dynamic conditions in the water, which introduce significant challenges to optical signal propagation. Turbulence, in particular, is caused by rapid yet spatially mild variations in the refractive index of water, primarily driven by fluctuations in temperature and salinity. Such turbulence distorts the optical wave, leading to signal degradation. Furthermore, scattering and absorption, which are also strongly affected by the characteristics of the underwater medium, are the primary contributors to the attenuation of optical waves in this environment \cite{Mobleyunderwaterlight, moreloceancolor, mobleylightandwater, quanindexrefraction}. These phenomena make reliable communication over long distances in underwater environments particularly challenging. Therefore, to effectively address these challenges, it is crucial to develop accurate analytical models that describe the spatial power spectrum under varying underwater conditions.

The majority of studies on laser beam propagation in turbulent optical channels have focused on single-wavelength transmission. To mitigate turbulence effects, wavelength diversity has been explored, leveraging the distinct propagation characteristics of different wavelengths to enhance UWOC performance \cite{kiasalehMWsc, SPIEMWShideh}. By using multiple wavelengths, the impact of scattering and absorption is mitigated, leading to more stable and robust links.
The theoretical foundations for turbulence modeling in UWOC have been well established in prior works \cite{SPIEMWShideh, andrews&phillipsbook, kiasalehgaussianMW}. This paper builds upon these models and applies them to evaluate the performance of multi-wavelength beam propagation under turbulence. Unlike prior studies that primarily focused on single-wavelength systems, this work systematically analyzes how wavelength diversity influences key metrics, such as the probability of fade, number of fades per second, and mean fade duration. The objective is to demonstrate that multi-wavelength transmission significantly improves link reliability, providing a performance-based evaluation of these established models and offering new insights into the benefits of wavelength diversity for UWOC.
%%%%%%%%%%%%%%%%%%%%%%%%%%%%%%%%%%%%%%%%%%%%%%%%%%%%%%%%%%%%%%%%%%%%%%%%%%%%%

\section{Channel and System Models}
\label{Channel and System Models}

In this study, we investigate the impact of multiple optical channels (N) on system performance. Each channel corresponds to an independent wavelength, and the received signal is a combination of these channels. The primary advantage of increasing N is that it provides diversity gain, which reduces the likelihood of deep fades by ensuring that at least one wavelength maintains an acceptable signal level. To model the UWOC system, we follow the best available literature, incorporating well-established techniques that account for factors such as absorption, scattering, and turbulence. This ensures that our approach is both accurate and relevant to current research in the field.

Turbulence causes an optical beam to undergo beam wander and spreading as a result of turbulent eddies. Additionally, the intensity of the optical signal exhibits random fluctuations due to scintillation. We first provide a concise overview of the power spectrum and scintillation index calculations for the UWOC. A specific ocean power spectrum model, known as the modified Nikishov spectrum, has been developed and covers most natural water conditions on Earth. Additionally, the log-normal distributed channel model is widely used to model weak turbulence regime. The system under analysis employs a Gaussian beam, an intensity-modulated/direct detection (IM/DD) technique, and on-off keying (OOK) as modulation scheme. A multiwavelength beam is investigated here for practical use, where beams of different wavelengths are combined at the transmitter and transmitted through a single optical path.

The performance of this proposed system, evaluated for a bit error rate of $10^{-9}$ without forward error correction (FEC), is analyzed using several metrics, including the probability of fade $(P_{FD})$, the average number of fades per second $(N_{FD})$, the mean fade duration $(T_{MFD})$, and the mean time between fades $(MTBF)$.

%%%%%%%%%%%%%%%%%%%%%%%%%%%%%%%%%%%%%%%%%%%%%%%%%%%%%%%%%%%%%%%%%%%%%%%%%%%%%

\subsection{Power Spectrum}
\label{Channel and System Models_A}

The power spectrum of the refractive index for oceanic water, based on the modified Nikishov spectrum, is given by \cite{nikishovunderwaterpowerspectrum}

\begin{equation}
    \begin{split}
    \Phi_n(\kappa) &= \frac{1}{4\pi} C_0 (\frac{\alpha^2 \chi_T}{\omega^2}) \epsilon^{-\frac{1}{3}} \kappa^{-\frac{11}{3}} \left[1 + C_1 (\kappa \eta)^{\frac{2}{3}}\right] \\
    &\quad \times \left[\omega^2 \exp \left(C_0 C_1^{-2} P_T^{-1} \delta\right) 
    + d_r \exp \left(C_0 C_1^{-2} P_S^{-1} \delta\right) \right. \\
    &\quad \left. - \omega (d_r + 1) \exp \left( \frac{-C_0 C_1^{-2} P_{TS}^{-1}}{2}\delta\right)\right],
    \end{split}
\end{equation}

where $\kappa$ denotes the spatial frequency magnitude, and $\varepsilon$ (measured in \text{m}$^2$/\text{s}$^3$) represents the energy dissipation rate. The constants $C_0$ and $C_1$ are assigned values of 0.72 and 2.35, respectively. Furthermore, $\chi_T$ corresponds to the thermal expansion coefficient of the Kolmogorov micro-scale length, while $\eta$ (expressed in \text{m}$^{-1}$) is given by $\eta = \nu^{\frac{4}{3}} \varepsilon^{-\frac{1}{4}}$, where $\nu$ is the kinematic viscosity.  

The parameter $\omega$, which represents the relative intensity of temperature and salinity fluctuations, is determined as $\omega = \frac{\alpha(dT/dz)}{\beta(dS/dz)}$, where $\alpha$ and $\beta$ denote the thermal expansion and saline contraction coefficients, respectively. Additionally, ${(dT/dz)}$ and ${(dS/dz)}$ correspond to the temperature and salinity gradients between the upper and lower boundaries.  

In Eq. (1), $P_T$ and $P_S$ denote the Prandtl numbers for temperature and salinity, respectively, while $P_{TS}$ is defined as half the harmonic mean of $P_T$ and $P_S$. The term $d_r$ represents the eddy diffusivity ratio, and $\delta$ is given by $\delta = 1.5c_1^2 (\kappa\eta)^{\frac{4}{3}} + C_1^3 (\kappa\eta)^2$ \cite{eddydiffusivity, yaounderwaterpowerspectrum}.  

%%%%%%%%%%%%%%%%%%%%%%%%%%%%%%%%%%%%%%%%%%%%%%%%%%%%%%%%%%%%%%%%%%%%%%%%%%%%%

\subsection{Scintillation index}
\label{Channel and System Models_B}

In the context of Rytov theory, numerous early studies on the statistical properties of an optical wave propagating through turbulence focused on the log-amplitude variance $\sigma_x^2(r, L)$. The scintillation index, which represents the normalized variance of the irradiance, is related to the log-amplitude variance as follows \cite{andrews&phillipsbook}

\begin{equation}
    \begin{aligned}
        \sigma_I^2(r, L) &= \frac{\langle I^2(r, L) \rangle}{\langle I(r, L) \rangle^2} - 1
    \end{aligned}
\end{equation}

With $I(r, L)$ denoting the intensity of the received signal, $r$ is the transverse distance from the center of the beam in the receiver aperture, $L$ is the propagation distance and $\left\langle \right\rangle$ denotes the ensemble average of the enclosed.

The scintillation index of a a multi-wavelength beam, as described by the developed power spectrum model \( \Phi_n(k) \), is given by the following expression \cite{SPIEMWShideh}

\begin{equation}
    \begin{aligned}
        \langle I^2(r, L) \rangle &= \sum\limits_{l=1}^{N} A_{l}^2(r, L) \Gamma_{l}^2(r, L) 
        \exp\left(4\sigma_{l}^2(r, L)\right) \\
        &\quad + 2 \sum\limits_{l_1=1}^{N} \sum\limits_{l_2=1}^{l_1-1} 
        A_{l_1}(r, L) A_{l_2}(r, L) \Gamma_{l_1 l_2}(r, L),
    \end{aligned}
\end{equation}
and 
\begin{equation}
    \langle I(r, L) \rangle^2 = \left[\sum\limits_{l=1}^{N} A_{l}(r, L) \Gamma_{l}(r,L)\right]^2.
\end{equation}

For the following equations, we define
\[
\begin{split}
\Gamma_l (r,L) &= \exp\left[ 2\left( \gamma_l (r,L) + \sigma_l^2 (r,L) \right) \right], \\
\Gamma_{l_1 l_2} (r,L) &= \Gamma_{l_1} \Gamma_{l_2} \exp\left[ 4 R_{l_1 l_2} (r,L; r,L) \right].
\end{split}
\]
Furthermore, for a Gaussian beam, \(\sigma_l^2(r, L)\), \(\gamma_l(r, L)\), and \(R_{l_1 l_2} (r,L; r,L)\) can be represented as follows

\begin{equation}
    \begin{split}
        \sigma_l^2\left(r, L\right)
        &= 2\pi^2 k_l^2 L \int_{0}^{1} \int_{0}^{\infty} \kappa \Phi_n (\kappa) \exp \left( \frac{-\kappa^2 \eta^2 L \Lambda_l}{ k_l} \right) \\
        &\quad \times I_0 \left\{ \left( 2 \kappa \Lambda_l \left| r \right| \right) - \left( \cos\left( \frac{\kappa^2 L \eta (1 - \eta) \hat{\Theta_l}}{k_l} \right) \right) \right\} \\
        &\quad d\kappa d\eta,
    \end{split}
\end{equation}

\begin{equation}
    \begin{split}
        \gamma_l\left(r, L\right) 
        &= -2\pi^2 k_l^2 L \int_{0}^{1} \int_{0}^{\infty} \kappa \Phi_n (\kappa) \\
        &\quad \times \left\{ 1 - \exp \left( \frac{-\kappa^2 \eta^2 L \Lambda_l}{k_l} \right) \right. \\
        &\quad \times \cos\left( \frac{\kappa^2 L \eta (1 - \eta) \hat{\Theta}_l}{k_l} \right)
        \left. \right\} d\kappa d\eta,
    \end{split}
\end{equation}
and
\begin{equation}
    \begin{split}
        R_{l_1l_2}\left(r, L; r, L\right) 
        &= 2\pi^2 k_{l_1} k_{l_2} L \int_{0}^{1} \int_{0}^{\infty} \kappa \Phi_n (\kappa) \\
        &\quad \times \text{Re} \left\{ J_0\left(\kappa \eta \left([\Theta_{l_1} - \Theta_{l_2}] - j(\Lambda_{l_1} + \Lambda_{l_2})\right) |r|\right) \right. \\
        &\quad \times \exp \left[ -j\kappa^2 L \left( \frac{(1-\eta) + \eta(\Theta_{l_1} - j\Lambda_{l_1})}{2k_{l_1}} \right. \right. \\
        &\quad \quad \left. \left. - \frac{(1-\eta) + \eta(\Theta_{l_2} + j\Lambda_{l_2})}{2k_{l_2}} \right) \eta \right] \\
        &\quad - J_0\left(\kappa \eta \left([\Theta_{l_1} - \Theta_{l_2}] - j(\Lambda_{l_1} - \Lambda_{l_2})\right) |r|\right) \\
        &\quad \times \exp \left[ -j\kappa^2 L \left( \frac{(1-\eta) + \eta(\Theta_{l_1} - j\Lambda_{l_1})}{2k_{l_1}} \right. \right. \\
        &\quad \quad \left. \left. + \frac{(1-\eta) + \eta(\Theta_{l_2} - j\Lambda_{l_2})}{2k_{l_2}} \right) \eta \right]
        \left. \right\} d\kappa d\eta.
    \end{split}
\end{equation}

respectively. For a detailed explanation of the calculation and parameters used, refer to \cite{SPIEMWShideh}.

%%%%%%%%%%%%%%%%%%%%%%%%%%%%%%%%%%%%%%%%%%%%%%%%%%%%%%%%%%%%%%%%%%%%%%%%%%%%%

\section{Channel's Time-Varying Properties}
\label{Channel's Time-Varying Properties}

The intensity function of the optical channel exhibits slow temporal variation, typically occurring on the order of milliseconds. This is in contrast to the bit durations associated with data rates in the range of several hundreds of megabit per second to a few gigabits per second. Consequently, the channel can be reasonably modeled as a slow-fading channel. The time-varying nature of the channel, however, is critical for calculating the overall bit error rate of the system as performance is significantly impacted by the level of the received signal. In such cases, understanding the parameters such as the fade duration and the frequency of fades are crucial. These parameters also play an important role in the performance analysis of packet-based communication systems. To better understand the temporal behavior of fades, it is necessary to analyze the temporal spectrum of irradiance fluctuations, represented here as $S_I(\omega)$. For a Gaussian beam, $S_I(\omega)$ is given by the expression \cite{andrews&phillipsbook}

\begin{equation}
    \begin{aligned}
        S_I(\omega) &= \frac{4.236 \sigma_I^2\left( r, L\right)}{\omega_t} 
        \int_0^1 \int_0^{\infty} \,
        e^{-\frac{\Lambda \omega^2 \zeta}{\omega_t^2}} t^{-1/2} 
        \left(t + \frac{\omega^2}{\omega_t^2}\right)^{-11/6} \\
        &\quad\mathrlap{\times}\quad e^{-\Lambda t \zeta} 
        \left\{ 1 - \cos\left[\left(t + \frac{\omega^2}{\omega_t^2}\right) \zeta (1 - \hat{\Theta} \zeta)\right] \right\} d t d\zeta.
    \end{aligned}
\end{equation}

Here, \( \sigma_I^2\left( r, L\right) \) represents the scintillation index, and \( \omega_t = \frac{V_T}{\sqrt{L / \kappa}} \), where \( V_T \) is the transverse wind velocity and  \( \sqrt{L / \kappa} \) indicates the size of the Fresnel zone.

The "width" (or root mean square bandwidth) for collimated Gaussian beam is defined by

\begin{equation}
    B_{\text{rms}} = \frac{1}{2 \pi} \left[ \frac{\int_0^{\infty} \omega^2 S_I(\omega) \, d\omega}{\int_0^{\infty} S_I(\omega) \, d\omega} \right]^{1/2}.
\end{equation}

Due to the infinite integration limit in (8), an alternative formula based on the definition of $B_{rms}$ can be used to calculate it as follows:

\begin{equation}
    B_{\text{rms}} = \frac{1}{2 \pi}  \left[- \frac{B_I^{{\prime\prime}}(0)}{B_I(0)} \right]^{1/2}.
\end{equation}

 Here, ${B_i}^{\prime\prime}(0) = \frac{\partial^2 {B_I}(\tau)}{\partial \tau^2} \Bigg|_{\tau = 0}$ with $B_I (\tau) = \frac{1}{2 \pi} \int_0^{\infty} S_I(\omega) \, \cos(\omega \tau) \, d\omega$ represents the temporal covariance function.
 For a Gaussian beam, $B_I (\tau)$ is expressed as

\begin{equation}
    \begin{aligned}
        B_I(\tau) &= 8 \pi^2 k^2 L \int_0^1 \int_0^{\infty} \, \kappa \Phi_n(\kappa) \\
        &\quad \times e^{-\frac{\Lambda L \zeta^2 \kappa^2}{k}} J_0\left(\kappa V_T \tau\right) \\
        &\quad \times \left\{1 - \cos \left[\frac{L\kappa^2}{k} \zeta(1 - \hat{\Theta} \zeta)\right]\right\} d \kappa d \zeta
    \end{aligned}
\end{equation}

where \( \Phi_n(\kappa)\) is the power spectrum, as defined in Section \ref{Channel and System Models_A}.

%%%%%%%%%%%%%%%%%%%%%%%%%%%%%%%%%%%%%%%%%%%%%%%%%%%%%%%%%%%%%%%%%%%%%%%%%%%%%

\section{Link Performance}
\label{Link Performance}

In conditions of weak turbulence, where scintillation can be modeled using a log-normal distribution, the probability of fade (PF) for a normalized threshold is given by 
\begin{equation}
    \begin{aligned}
        F_{th} &= 10 \log_{10}\left(\frac{\overline{I_p(0)}}{I_{th}}\right) \, \text{(in dB)}
    \end{aligned}
\end{equation}

with \( I_{th} \) representing the fade threshold level below the mean on-axis irradiance, and $\overline{I_p(0)}$is the mean intensity on the optical axis, which is identical to the peak intensity of the beam in the case of a Gaussian beam.

The on-axis probability of fade $P_{\text{FD}}$ for a finite aperture diameter, where $n = 1, \dots, N$ denotes the number of distinct wavelengths, is expressed as \cite{kiasalehpfd}

\begin{equation}
    \begin{aligned}
        P_{FD,\,n}^{\text{on-axis}} &= p\left\{ I_p(r) \leq I_{th} \right\}
        = \int_0^{I_{th}} f_{I_p}(x) \, dx \\
        &= \frac{1}{2} \left\{ 1 + \operatorname{erf} \left[ \frac{\frac{\sigma_I^2\left( r, L\right)}{2} - 0.23 F_{th} }{\sqrt{2} \sigma_I (r,L)} \right] \right\}.
    \end{aligned}
\end{equation}

For an N-channel system, the overall probability of fade is determined using

\begin{equation}
    P_{\text{FD}}^{\text{on-axis}} = \prod_{n=1}^{N} P_{\text{FD,\,n}}^{\text{on-axis}}
\end{equation}

where $P_{\text{FD,\,n}}$ represents the probability of fade for the $n$th wavelength. As N increases, the combined probability of fading decreases, thereby improving link stability. This effect is further validated in Section V, where numerical results demonstrate that higher N values lead to lower fade probabilities and longer mean time between fades.

The on-axis average number of fades per second (FPS) for a Gaussian beam is given by \cite{kiasalehpfd}

\begin{equation}
    N_{F D}^{\text{on-axis}} = B_{\text{rms}} \exp \left\{-\frac{\left[0.5 \sigma_I^2(r, L) - 0.23 F_{th}\right]^2}{2 \sigma_I^2(r,L)}\right\}.
\end{equation}

The mean fade duration (MFD) is given by \cite{kiasalehpfd}

\begin{equation}
    T_{M F D} = \left\{\frac{P_{FD}^{\text{on-axis}}}{N_{F D}^{\text{on-axis}}}\right\}.
\end{equation}

Once can define $T_{MTBF}$ as the mean time between fades (MTBF), which is expressed as \cite{kiasalehpfd}

\begin{equation}
    T_{M T B F} = \left\{\frac{(1 - P_{FD}^{\text{on-axis}})}{N_{F D}^{\text{on-axis}}}\right\}.
\end{equation}

%%%%%%%%%%%%%%%%%%%%%%%%%%%%%%%%%%%%%%%%%%%%%%%%%%%%%%%%%%%%%%%%%%%%%%%%%%%%%

\section{Numerical Results}
\label{Numerical Results}

The results presented in this section provide simulation-based insights into the effect of wavelength diversity on underwater optical communication. While this study does not incorporate experimental validation, the trends observed are consistent with existing experimental studies on turbulence-induced fading in underwater environments. Future work should focus on empirical validation using controlled underwater testbeds, where multi-wavelength transmission can be tested under varying turbulence conditions.

The numerical analysis is performed with several parameters fixed for generality. Specifically, the temperature is maintained at $20^{\circ}\mathrm{C}$, salinity at $35 \text{ppt}$, and pressure at $0 \text{dBar}$, with $\omega = -0.3508$, $\varepsilon = 10^{-2} m^2/s^3$ and $\chi_T = 10^{-5} K^2/S$. A collimated beam is assumed ($F_0 = \infty$), and the scintillation index is calculated at the beam's focal point ($r = 0$). Since absorption plays a dominant role in underwater environments, the blue region of the visible spectrum (450–485 nm) experiences the least attenuation compared to other spectral regions. To highlight the benefits of wavelength diversity, multiple wavelengths between 480 and 600 nm are analyzed, covering the blue-green, green, and yellow regions of the spectrum. The fade threshold $F_{th}$ is considered to be 5 dB, which is then employed to determine the probability of fade and other relevant parameters. The results presented in the figures provide significant insights into the performance of underwater optical communication systems under varying conditions.

Figure 1 illustrates the probability of fading ($P_\text{FD}$) as a function of propagation distance ($L$) for three different wavelengths: $\lambda = 480 \, \text{nm}$, $\lambda = 532 \, \text{nm}$, and $\lambda = 600 \, \text{nm}$. The behavior of $P_\text{FD}$ clearly demonstrates the impact of wavelength on channel performance. At shorter wavelengths ($\lambda = 480 \, \text{nm}$), the probability of fading is higher compared to longer wavelengths, primarily due to the stronger scattering experienced by shorter wavelengths in underwater environments. In contrast, longer wavelengths, such as $\lambda = 600 \, \text{nm}$, exhibit a lower fading probability, reflecting greater resilience to turbulence and scattering. When comparing the results for the fade threshold of 3 dB in Figure 1(a) and 5 dB in Figure 1(b), the probability of fading at 3 dB is higher, indicating that the system is more susceptible to fading under the lower threshold. Specifically, for the 3 dB threshold, the probability of fading reaches approximately 0.42 at 15 m, while for the 5 dB threshold, it is around 0.25. The system's vulnerability to fading is greater at the lower threshold, highlighting the increased sensitivity to minor signal fluctuations.

\begin{figure}
    \centering
    \begin{tabular}{c}
        \includegraphics[width=3.7in]{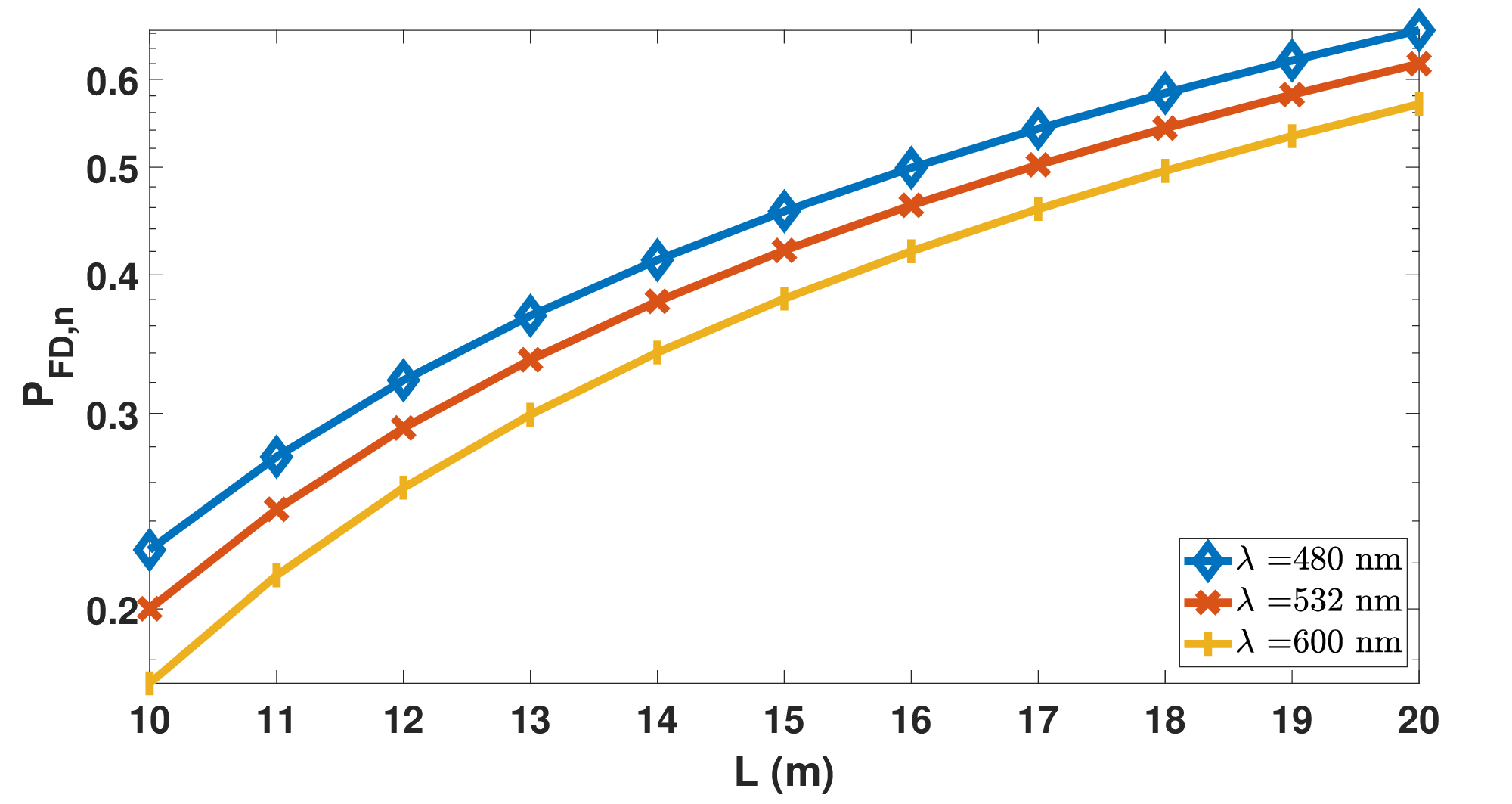} \\
        \small (a) 3 dB \\
    \end{tabular}
    \vspace{0.35cm}
    \begin{tabular}{c}
        \includegraphics[width=3.7in]{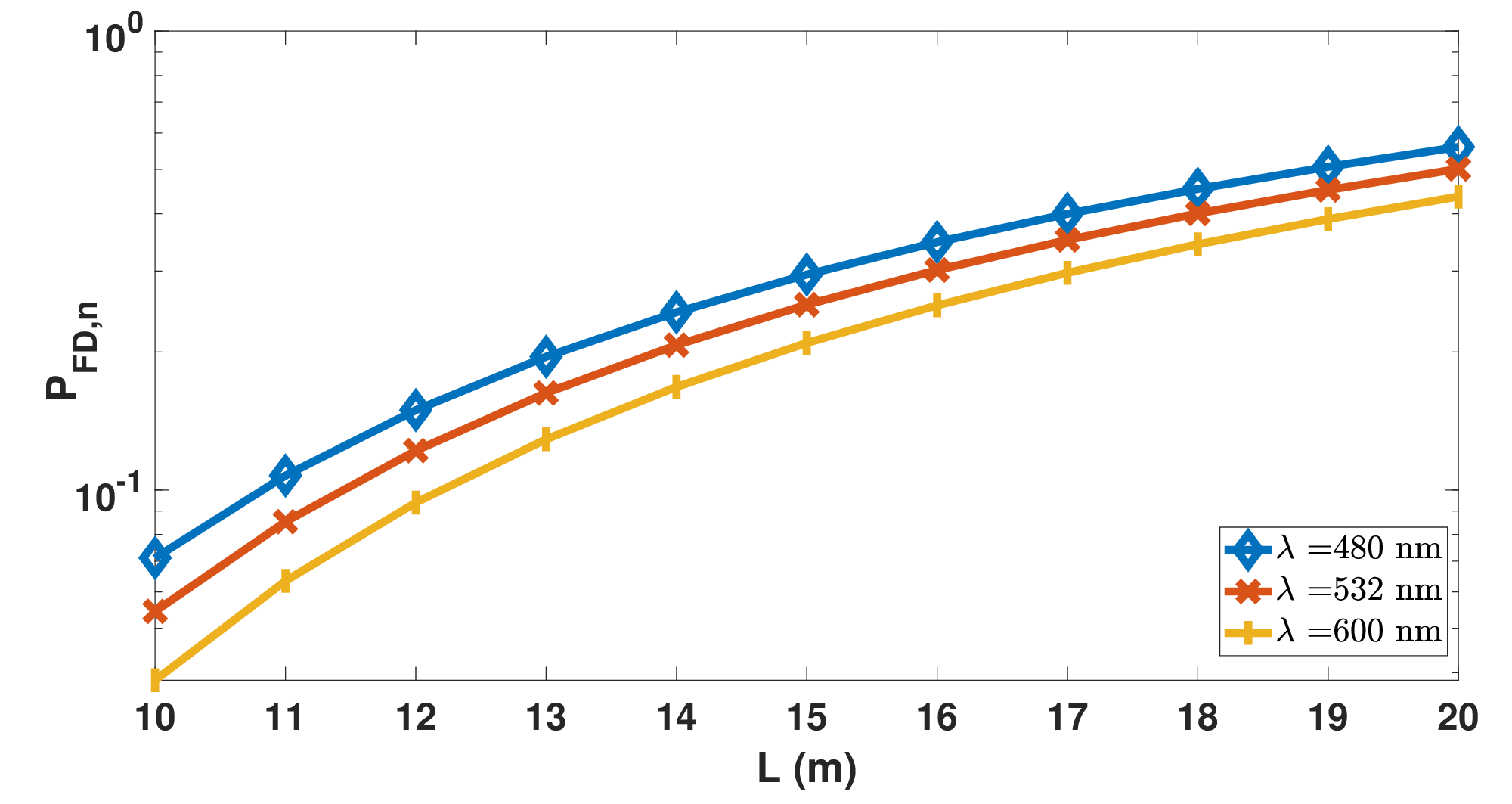} \\
        \small (b) 5 dB \\
    \end{tabular}
    \caption{Probability of fading vs. Propagation distance for one wavelength}
    \label{figure1}
\end{figure}

In Figure 2, the influence of diversity is further highlighted by showing the probability of fading as a function of propagation distance, utilizing multi-wavelength system with one, two, and three wavelengths. For $N=2$, pairs of wavelengths ($\lambda = 480$ nm and $\lambda = 532$ nm), ($\lambda = 480$ nm and $\lambda = 600$ nm) and ($\lambda = 532$ nm and $\lambda = 600$ nm) are used to demonstrate wavelength diversity. For $N=3$, three wavelengths ($\lambda = 480$ nm, $\lambda = 532$ nm, and $\lambda = 600$ nm) are employed to further enhance diversity. The chosen wavelength range (480-600 nm) is limited by the absorption characteristics of underwater environments.

For pairs of wavelengths (\(N = 2\)), the probability of fading increases with distance, reflecting the detrimental effects of turbulence and fading. However, as the number of wavelengths increases to \(N = 3\), there is a substantial improvement in reliability, attributed to the diversity gain provided by multiple links. In Figure 2(b), for \(N = 3\), the system achieves lower fading even at longer distances, demonstrating the robustness of link diversity in mitigating fading. With a fade threshold of 3 dB in Figure 2(a), the probability of fading increases exponentially, from 0.007 at 10 m to 0.2 at 20 m, compared to a less pronounced drop at the 5 dB threshold in Figure 2(b), where the probability is \(10^{-4}\) at 10 m and \(10^{-1}\) at 20 m for \(N = 3\). The diversity gain is more prominent at the lower threshold, where the probabilities for \(N = 2\) and \(N = 3\) increase more significantly, emphasizing the crucial role of increasing the number of wavelengths to counteract the effects of fading.

\begin{figure}
    \centering
    \begin{tabular}{c}
        \includegraphics[width=3.7in]{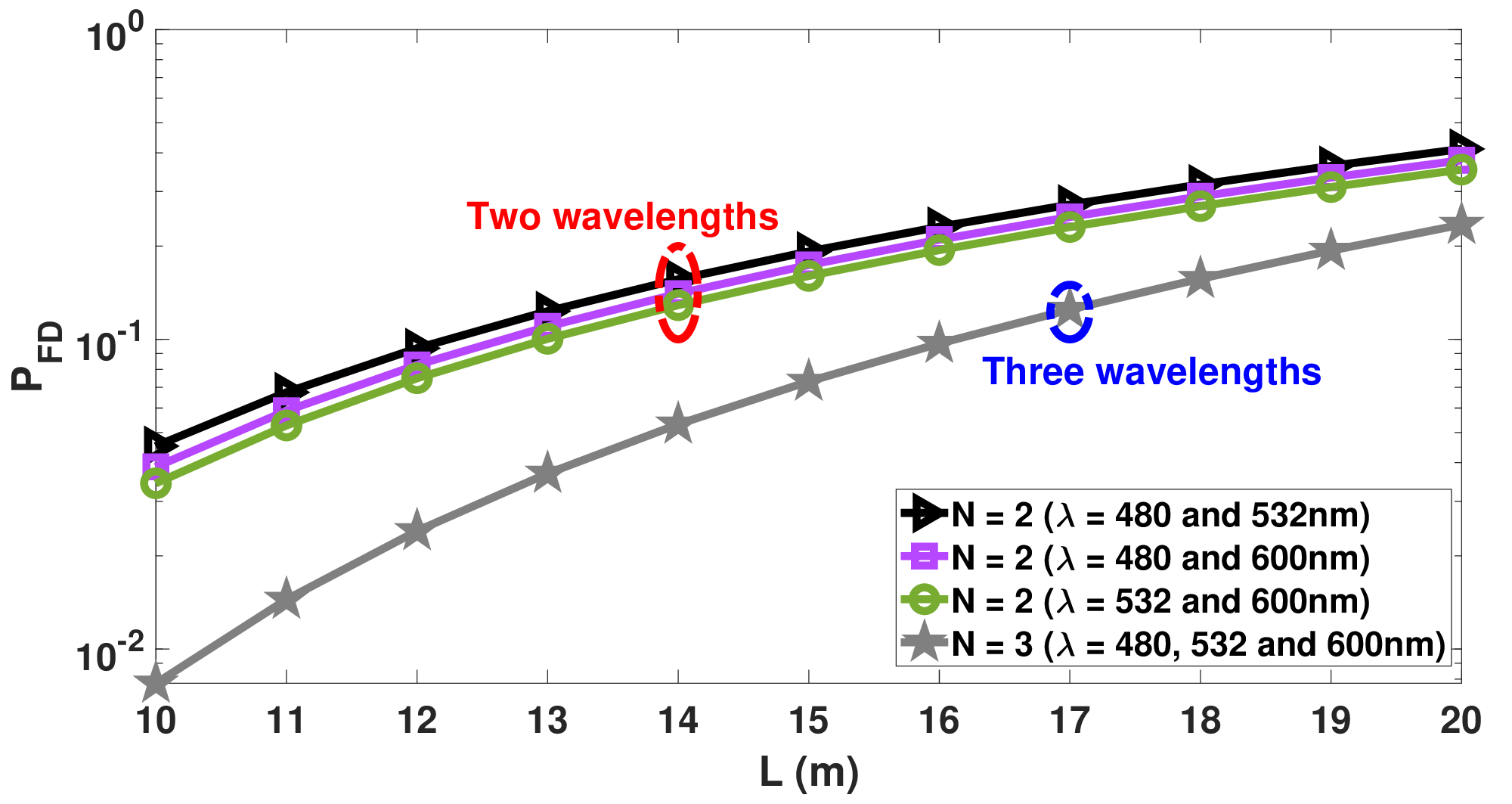} \\
        \small (a) 3 dB \\
    \end{tabular}
    \vspace{0.35cm}
    \begin{tabular}{c}
        \includegraphics[width=3.7in]{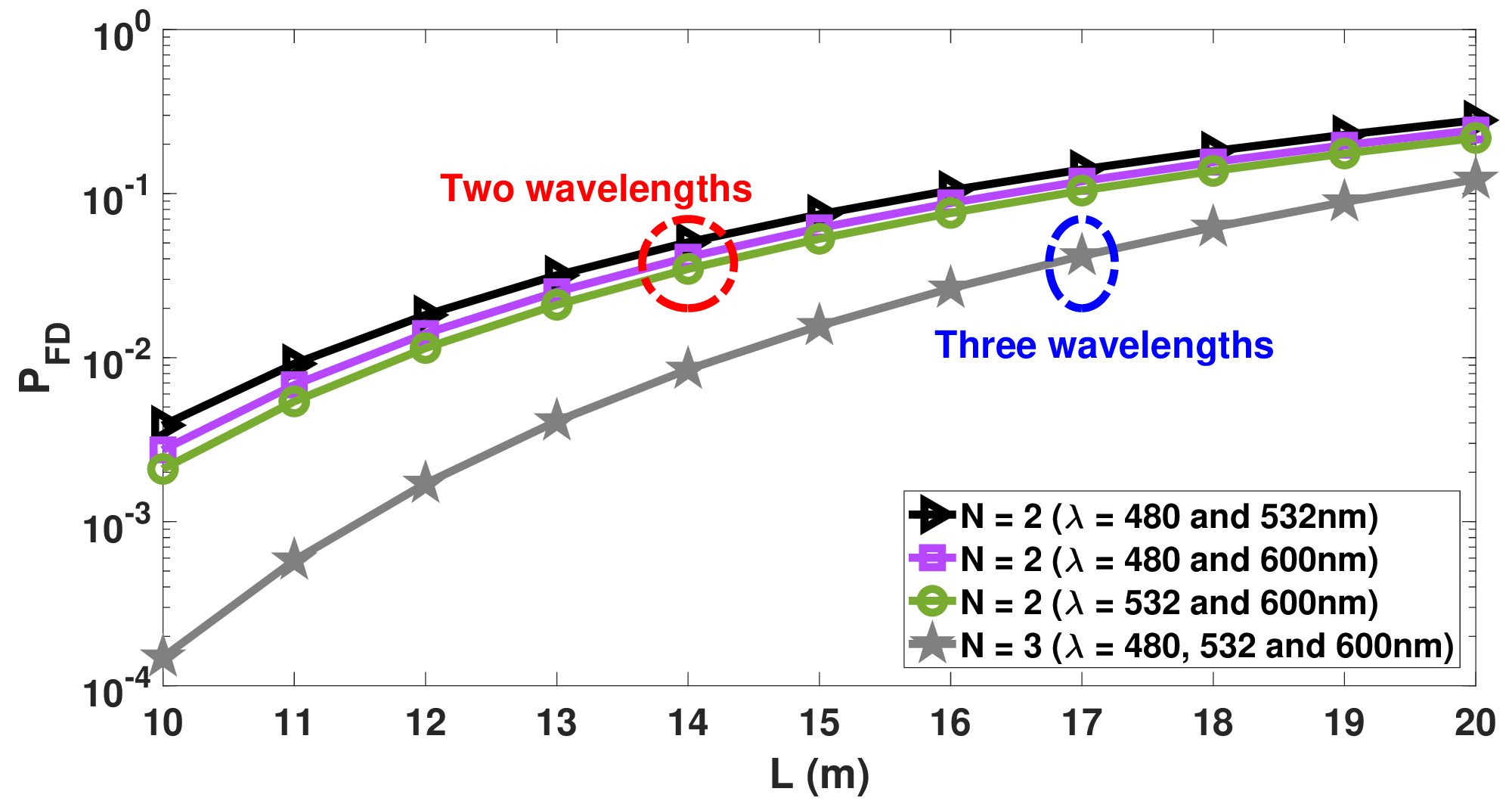} \\
        \small (b) 5 dB \\
    \end{tabular}
    \caption{Probability of fading vs. Propagation distance for multi-wavelength}
    \label{figure2}
\end{figure}

Figure 3 shows the average number of fades per second (\(N_\text{FD}\)) as a function of the propagation distance \(L\). To compute \(N_\text{FD}\), we first determine \(\sigma_I^2(r, L)\) using Equations (2), (3), and (4). The results indicate that as \(L\) increases, \(N_\text{FD}\) grows exponentially, highlighting the increasing likelihood of fade detection over longer propagation paths. This behavior is driven by the accumulation of turbulence-induced scintillation effects, as described by \(\sigma_I^2(r, L)\). 

As shown in Figure 3(a), shorter wavelengths, such as \(\lambda = 480 \, \text{nm}\), result in higher values of \(N_\text{FD}\) due to stronger scattering interactions, whereas longer wavelengths, such as \(\lambda = 600 \, \text{nm}\), exhibit lower values, indicating reduced sensitivity to turbulence. Beyond \(L \approx 17 \, \text{m}\), the growth of \(N_\text{FD}\) slows, suggesting a saturation effect as the contribution of turbulence stabilizes.  Figure 3(b) further emphasizes the dependence of \(N_\text{FD}\) on propagation distance and wavelength. The trends observed remain consistent when employing \(N = 2\) and \(N = 3\), demonstrating significant improvements as \(N_\text{FD}\) decreases sharply compared to \(N = 1\). For instance, at \(L = 17\) m, applying \(N = 2\) and \(N = 3\) results in approximately 58\% and 89\% improvement, respectively.

\begin{figure}
    \centering
    \begin{tabular}{c}
        \includegraphics[width=3.7in]{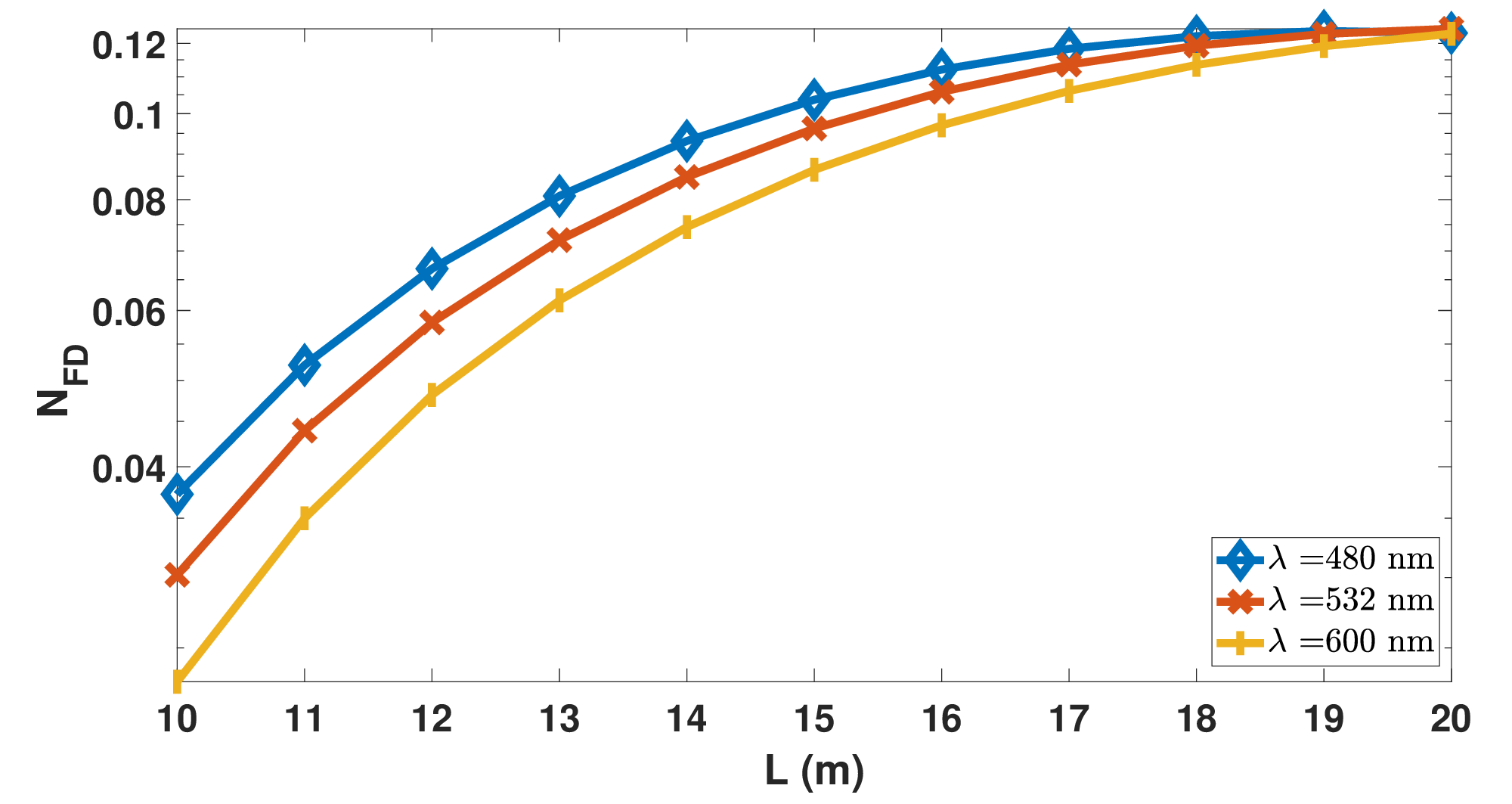} \\
        \small (a) One wavelength \\
    \end{tabular}
    \vspace{0.35cm}
    \begin{tabular}{c}
        \includegraphics[width=3.7in]{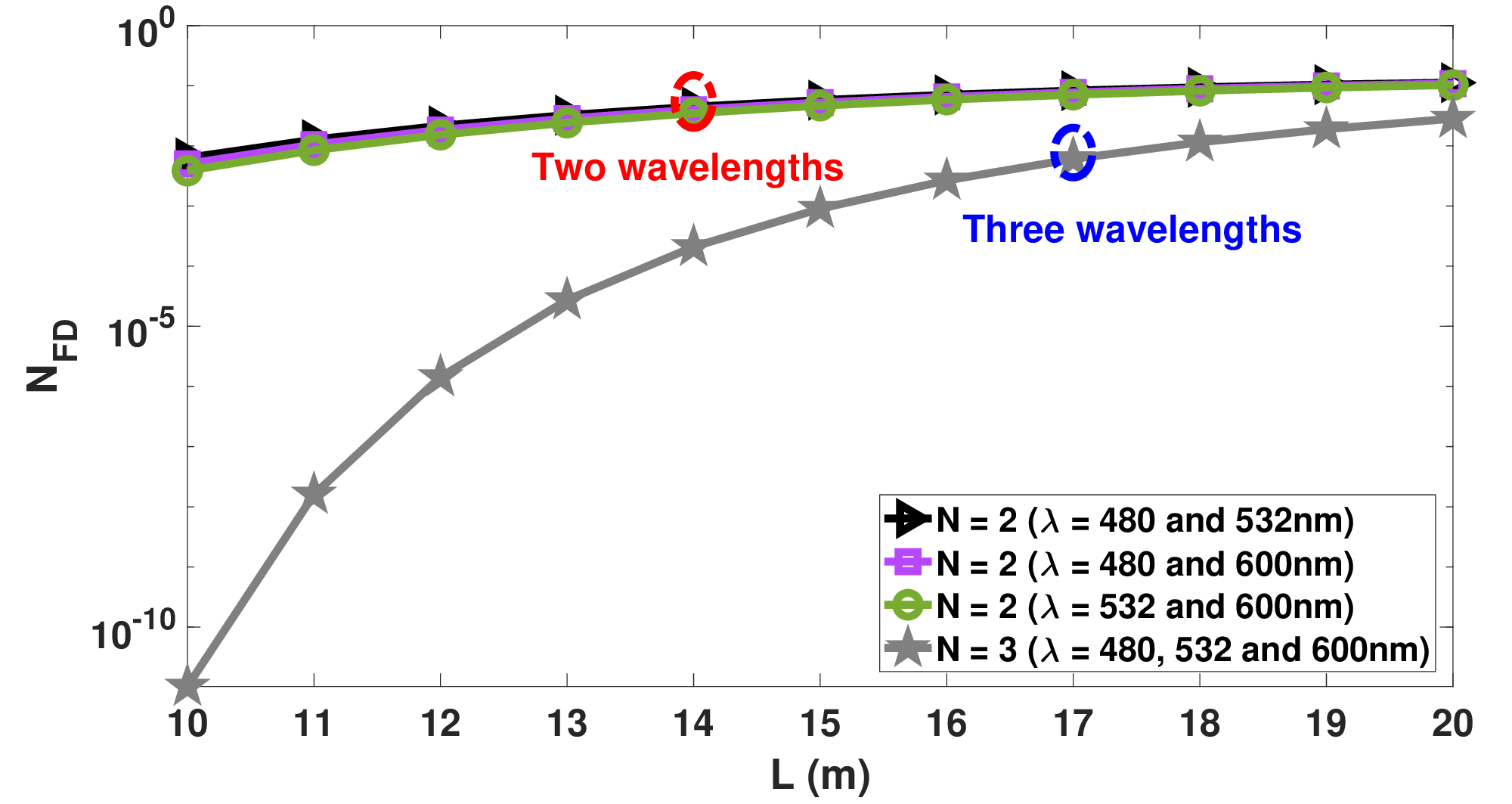} \\
        \small (b) Multi-wavelength \\
    \end{tabular}
    \caption{$N_{FD}$ vs. Propagation distance for different wavelengths}
    \label{figure3}
\end{figure}

Figure 4 shows the mean fade duration (MFD) as a function of the propagation distance. In Figure 4(a), as the propagation distance increases, the MFD also rises, with shorter wavelengths experiencing a higher likelihood of fading compared to longer wavelengths. In Figure 4(b), when \( N = 2 \), the MFD is reduced, highlighting the effectiveness of using multiple wavelengths to improve system performance in the presence of turbulence. Increasing \( N \) to 3 leads to an even more significant decrease in MFD. For instance, at \( L = 13 \, \text{m} \), the MFD is around 2.2 for a single wavelength, but it drops to 0.8 for \( N = 2 \) and further decreases to \( 6 \times 10^{-3} \) for \( N = 3 \).

Reducing the MFD enhances system performance by improving signal reception reliability. Shorter fade durations lead to fewer deep fades, resulting in a higher signal to noise ratio (SNR) and reduced bit error rates (BER). This, in turn, strengthens the communication link. Additionally, using multiple wavelengths further minimizes prolonged fades, boosting throughput and system capacity in the presence of turbulence and attenuation

\begin{figure}
    \centering
    \begin{tabular}{c}
        \includegraphics[width=3.7in]{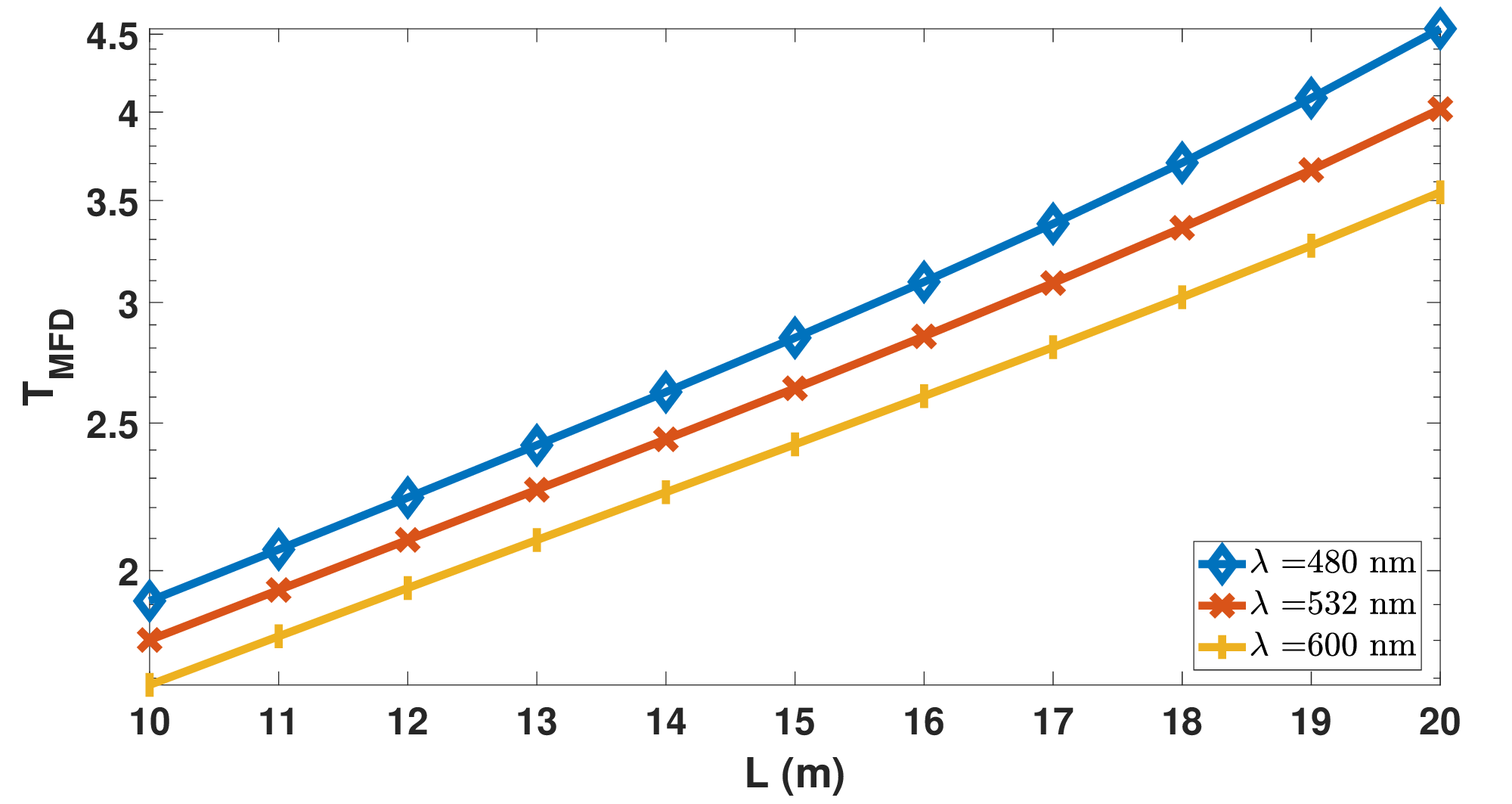} \\
        \small (a) One wavelength \\
    \end{tabular}
    \vspace{0.35cm}
    \begin{tabular}{c}
        \includegraphics[width=3.7in]{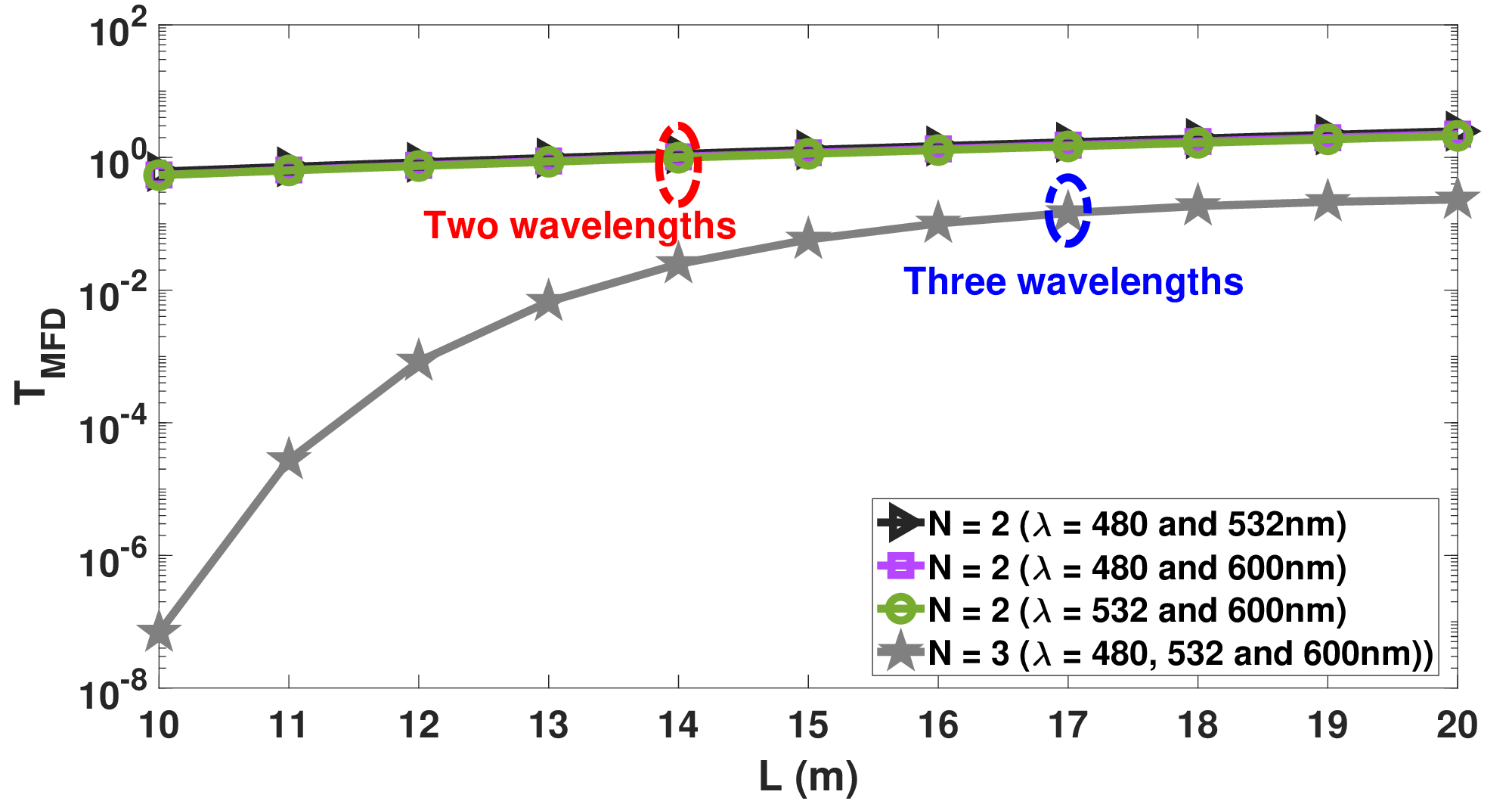} \\
        \small (b) Multi-wavelength \\
    \end{tabular}
    \caption{$T_{MFD}$ vs. Propagation distance for different wavelengths}
    \label{figure4}
\end{figure}

Figure 5 shows the mean time between fades (MTBF), revealing a noticeable decrease as the propagation distance grows. In Figure 5(a), it is seen that shorter wavelengths generally result in lower MTBF at longer distances. In Figure 5(b), the MTBF decreases significantly with increasing $N$ particularly at shorter distances, indicating the advantage of using multiple wavelengths in improving the system’s resilience to fading.

MTBF is a key performance metric that reflects signal reliability and continuity. A higher MTBF means fewer fading events, reducing disruptions and maintaining a stable connection. This is particularly crucial for minimizing deep fades that degrade signal quality. Multi-wavelength systems (\(N > 1\)) further enhance MTBF by distributing fading effects across different wavelengths, improving link robustness, increasing availability, and lowering BER.

\begin{figure}
    \centering
    \begin{tabular}{c}
        \includegraphics[width=3.7in]{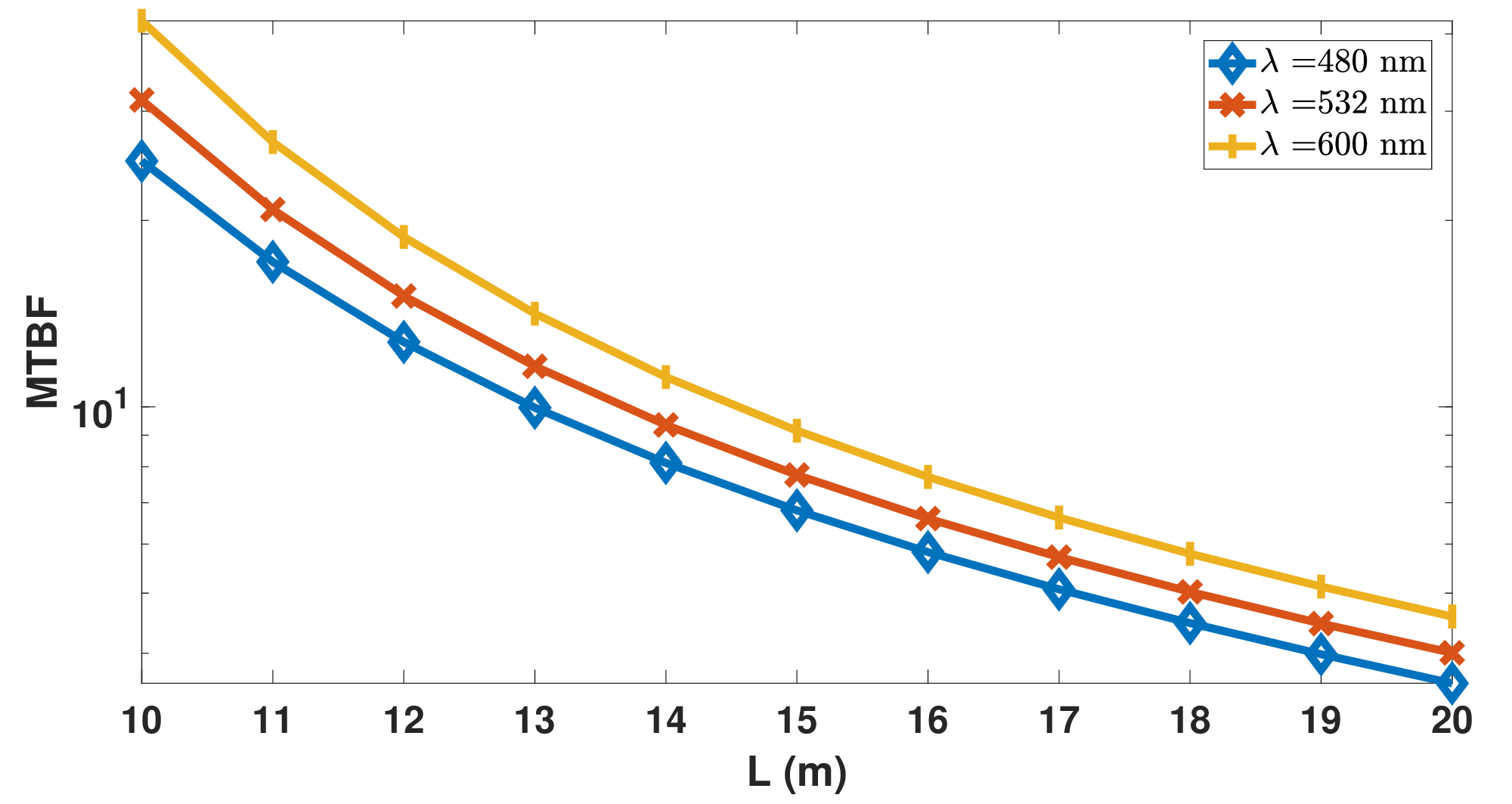} \\
        \small (a) One wavelength \\
    \end{tabular}
    \vspace{0.35cm}
    \begin{tabular}{c}
        \includegraphics[width=3.7in]{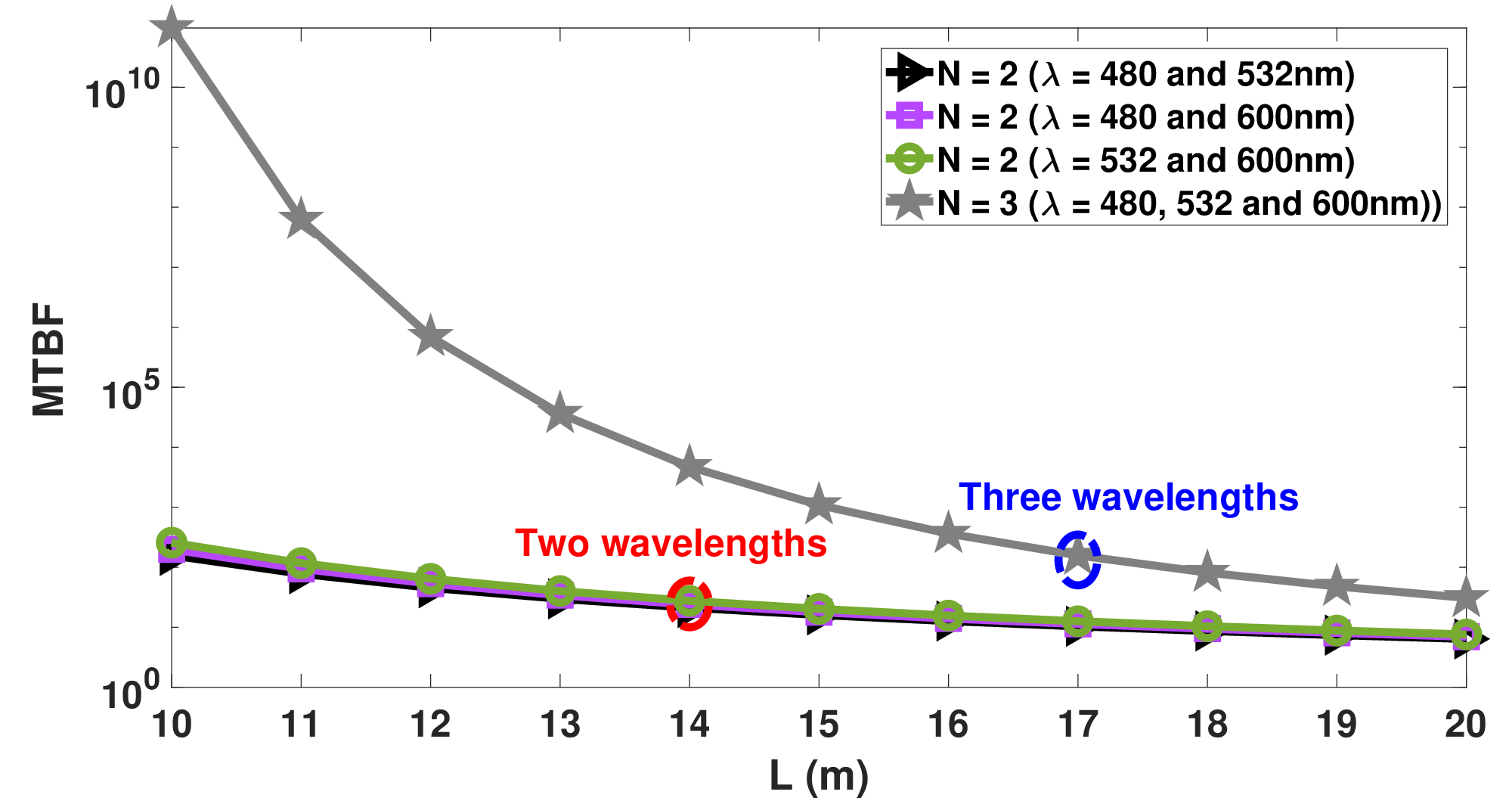} \\
        \small (b) Multi-wavelength \\
    \end{tabular}
    \caption{MTBF vs. Propagation distance for different wavelengths}
    \label{figure5}
\end{figure}

%%%%%%%%%%%%%%%%%%%%%%%%%%%%%%%%%%%%%%%%%%%%%%%%%%%%%%%%%%%%%%%%%%%%%%%%%%%%%

\section{Conclusion}
\label{Conclusion}

In this paper, we examined the impact of wavelength diversity on UWOC performance, utilizing previously established turbulence models. Our findings highlight that increasing the number of independent optical channels (N) significantly improves link performance by reducing the probability of fade and increasing the mean time between fades. These results emphasize the practical benefits of multi-wavelength transmission in mitigating turbulence-induced signal degradation.

The results highlight the role of wavelength selection and link diversity in enhancing communication reliability. Longer wavelengths reduce fading probability ($P_\text{FD}$), minimizing signal degradation due to turbulence. The analysis of fades per second ($N_\text{FD}$) shows that multiple wavelengths reduce fading events, improving signal stability. Wavelength diversity also lowers mean fade duration (MFD), ensuring shorter signal interruptions. A higher mean time between fades (MTBF) further supports this by extending the duration between fading events, leading to more consistent signal reception.
While a lower fade threshold increases sensitivity to fading, wavelength diversity mitigates this effect, ensuring reliable performance over longer distances. Additional improvements can be made by optimizing receiver sensitivity, increasing transmitter power, or combining both strategies to raise the fade threshold. Together with wavelength diversity, these enhancements help overcome the challenges of underwater optical communication, contributing to more robust and efficient systems for underwater applications.

\bibliographystyle{IEEEtran}

\bibliography{IEEEabrv,IEEEexample}

@article{korotkovalightinturbulentocean,
	author = "O. Korotkova", 
	title  = "Light propagation in a turbulent ocean", 
	journal= "Elsevier",  
	volume = "64", 
	pages  = "1-43",
	year   = "2018",
	doi = "10.1016/bs.po.2018.09.001",
	note   = "[doi:10.1016/bs.po.2018.09.001]"
}

@article{Mobleyunderwaterlight, 
	author = "C. D. Mobley", 
	title  = "Comparison of numerical models for computing underwater light fields", 
	journal= "Appl. Opt.",  
	volume = "32", 
	number = "36",
	pages  = "7484-7504",
	year   = "1993"
}

@book{mobleylightandwater, 
	author = "C. D. Mobley", 
	title  = "Light and Water: Radiative Transfer in Natural Waters", 
	publisher= "Academic",  
	address = "San Diego, CA, USA",
	year   = "1994"
}

@article{moreloceancolor, 
	author = "A. Morel and L. Prieur", 
	title  = "Analysis of variations in ocean color", 
	journal= "Limnol. Oceanogr.",  
	volume = "22", 
	pages  = "709-722",
	year   = "1977"
}

@article{quanindexrefraction, 
	author = "X. Quan and E. S. Fry", 
	title  = "Empirical equation for the index of refraction of seawater", 
	journal= "Appl. Opt.",  
	volume = "34", 
	pages  = "3477-3480",
	year   = "1995"
}

@article{kiasalehMWsc, 
	author = "K. Kiasaleh", 
	title  = "Scintillation index of a multiwavelength beam in turbulent atmosphere", 
	journal= "J. Opt. Soc. Am. A",  
	volume = "21", 
	pages  = "1452-1454",
	year   = "2004"
}

@article{yaounderwaterpowerspectrum,
    author  = {J. Yao and M. Elamassie and O. Korotkova},
    title   = {Spatial power spectrum of natural water turbulence with any average temperature, salinity concentration, and light wavelength},
    journal = {J. Opt. Soc. Amer. A},
    volume  = {37},
    pages   = {1614--1621},
    year    = {2020}
}

@article{nikishovunderwaterpowerspectrum,
    author  = {M. Elamassie and M. Uysal and Y. Baykal and M. Abdallah and K. Qaraqe},
    title   = {Effect of eddy diffusivity ratio on underwater optical scintillation index},
    journal = {Journal of the Optical Society of
    America A,},
    volume  = {34},
    number  = {11},
    pages   = {1969--1973},
    year    = {2017}
}

@article{eddydiffusivity,
    author  = {P. R. Jackson and C. R. Rehmann},
    title   = {Laboratory measurements of differential diffusion in a diffusively stable, turbulent flow},
    journal = {J. Phys. Oceanogr.},
    volume  = {33},
    number  = {8},
    pages   = {1592--1603},
    year    = {2003}
}

@book{andrews&phillipsbook, 
    author    = "L. C. Andrews and R. L. Phillips", 
    title     = "Laser Beam Propagation through Random Media", 
    edition   = "2nd", 
    publisher = "SPIE",  
    year      = "2005"
}

@article{kiasalehgaussianMW, 
	author = "K. Kiasaleh", 
	title  = "On the scintillation index of a multiwavelength Gaussian beam in a turbulent free-space optical communications channel", 
	journal= "J. Opt. Soc. Am. A",  
	volume = "23", 
	pages  = "557-566",
	year   = "2006"
}

@article{SPIEMWShideh,
    author = "Shideh Tayebnaimi and Kamran Kiasaleh", 
    title = "Scintillation index analysis for multi-wavelength Gaussian beams in turbulent underwater channels", 
    journal = "Optical Engineering", 
    volume = "63", 
    number = "12", 
    pages = "128105", 
    year = "2024", 
    doi = "10.1117/1.OE.63.12.128105"
}

@article{kiasalehpfd,
	author = "K. Kiasaleh", 
	title  = "Performance analysis of free-space on-off keying optical communication systems impaired by turbulence", 
	journal = "Proc. SPIE", 
	volume = "4635", 
	pages = "150-161", 
	year = "2002"
}

\end{document}